\documentclass[reprint, amsmath, amssymb, aps, prb]{revtex4-1}
\usepackage[utf8]{inputenc}
\usepackage{graphicx}
\usepackage{xcolor}
\usepackage{braket}
\usepackage{array}
\newcolumntype{C}[1]{>{\centering\let\newline\\\arraybackslash\hspace{0pt}}m{#1}}

\newcommand*{\W}{\mathcal{W}}
\newcommand*{\tr}[1]{\mathrm{tr}\big(#1\big)}

\begin{document}

\title{Interplay between long-range hoppings and disorder in topological systems}

\author{Beatriz P\'erez-Gonz\'alez}
\email{bperez03@ucm.es} 
\author{Miguel Bello}
\email{miguel.bello@icmm.csic.es} 
\author{\'Alvaro G\'omez-Le\'on} 
\author{Gloria Platero} 
\affiliation{Instituto de Ciencia de Materiales de Madrid (ICMM-CSIC)}
\affiliation{Materials Science Factory, Instituto de Ciencia de Materiales de Madrid, CSIC}
\date{\today}

\begin{abstract}
  We extend the standard SSH model to include long-range hopping amplitudes and disorder, and
  analyze how the electronic and topological properties are affected. We show that
  long-range hoppings can change the symmetry class and the topological
  invariant, while diagonal and off-diagonal disorder lead to Anderson
  localization. Interestingly we find that the Lyapunov exponent $\gamma(E)$ can
  be linked in two ways to the topological properties in the presence of
  disorder: Either due to the different response of mid-gap states to increasing
  disorder, or due to an extra contribution to $\gamma$ due to the presence of
  edge modes. Finally we discuss its implications in realistic transport
  measurements.
\end{abstract}


\maketitle

\section{Introduction}
The dimer chain is a canonical model in condensed matter, widely used to study
topological properties in atomic chains with staggered hopping amplitudes
\cite{Peierls1955,Jackiv1976,SSH1979,ssh1980}. For the case of just nearest-neighbor hoppings (NN) it reduces to the well-known SSH model \cite{ssh1980}. However, the addition of long-range hopping amplitudes is important, as they are ubiquitous in real systems and their role can be crucial \cite{metallic2vec, viyuela2018,
diptiman1}.  The effect of disorder in quantum systems can also be significant
\cite{AndersonLocalization}. In topological systems in particular, its study has
been focused on the change of the topological invariants
\cite{TAI1,TAI2,TAI3,Vozmediano}.  However, its relation with localization has
been typically overlooked, with some exceptions \cite{Disorder-QW,
Altland2015,quasiperiodicSSH, dutta}.   

In this work we discuss the interplay between long-range hopping amplitudes and disorder in
topological chains. We show that long-range hoppings connecting sites within the
same sublattice break chiral symmetry and changes the topological phase,
leading to edge modes without the original topological protection. For large
enough hopping amplitudes, the edge states mix with the bulk bands and the system becomes metallic. In
contrast, long-range hopping amplitudes connecting different sublattice sites preserve
chiral symmetry and allow to increase the value of the topological invariant
and the number of edge modes.  

In the presence of disorder Anderson localization happens. However, differences
between diagonal (DD) and off-diagonal disorder (ODD) can be observed, when just
NN hopping amplitudes are present due to the breaking of chiral symmetry in the first case
but not the second. Furthermore, the localization properties for mid-gap and
bulk states are different, and the Lyapunov exponent exhibits an extra
contribution due to edge modes. This makes topology and localization intertwine
in an interesting way.  When next-nearest-neighbors (NNN) hopping amplitudes are included, the effect of both types of disorder is similar due to the lack of chiral symmetry in both cases. However, it is possible to find signatures of topology in transport measurements, even deep within the metallic phase and for weak disorder.  

\section{Model}
The standard SSH model is described in terms of non-interacting, spinless
electrons populating a chain with alternating hopping amplitudes between neighboring 
sites. Its generalized Hamiltonian is: 
\begin{equation}
  H_R = \sum_{|i-j|\leq R} (J_{i,j}+\epsilon_{i,j})c^\dagger_i c_j
  + \mathrm{H.c.} \,,
\label{eq:hdimer}
\end{equation}
where $c^\dagger_i$ creates a fermion at the $i$th site and $J_{i,j}$, $i\neq
j$, is the hopping amplitude connecting the $i$th and the $j$th sites. $R$ is
the maximum range of the hopping and we set $J_{i,i}=0$.  Diagonal and
off-diagonal disorder are introduced through $\epsilon_{i,j}$, for $i=j$ and $i
\neq j$, respectively. For the numerical results we have considered
$\epsilon_{i,j}\in[-w/2,w/2]$ homogeneously distributed, although other choices
are possible \cite{odagaki2,LogOffdiag1,soukoulis81}. Because $J_{ij}$ are
functions of the distance $n=|i-j|$, to simplify our notation we separate
hopping processes connecting sites within the same sublattice $J_{i,i\pm
n}\equiv J_n$, with $n$ even, and hopping processes connecting sites in
different sublattices $J_{2i-n,2i}\equiv J_n$, and $J_{2i+n,2i}\equiv J'_n$,
with $n$ odd. 

In absence of disorder, the Hamiltonian can be diagonalized in $k$-space and
written in terms of the Pauli matrices as
$\mathcal{H}_R=d_{0}(k)\mathbf{1}+\vec{d}(k)\cdot\vec{\sigma}$, with 
\begin{gather}
	d_0(k) = \sum_p 2J_{2p}\cos(pk) \,, \quad d_z(k) = 0 \,, \\
  d_x(k) = \sum_p \left\{J'_{2p-1}\cos[(p-1)k] + J_{2p-1}\cos(pk)\right\} \,, 
    \\
	d_y(k) = \sum_p \left\{J_{2p-1}\sin(pk) - J'_{2p-1}\sin[(p-1)k]\right\} \,,
\end{gather}
where $\mathcal{H}_R$ acts on the pseudospinor ${\Psi_{k}=(a_{k},b_{k})^{T}}$ and $p$
ranges from 1 to $\lfloor(R+1)/2\rfloor$ ($\lfloor \cdots \rfloor$ denotes the
floor function). The dispersion relation is $E_\pm(k)=d_0(k)\pm|\vec{d}(k)|$,
where ``$+$'' and ``$-$'' correspond to the conduction and valence band,
respectively (see Fig.~\ref{fig:2vec_spectrum_loclength},(b)). 

\section{Absence of disorder}
The topological properties of the standard SSH model are captured with
the Zak phase \cite{zakphase1989}, or equivalently the winding number $\W$ of
the Bloch vector $\vec{d}(k)$ \cite{delplace2011}.  

As the SSH model has time-reversal, particle-hole and chiral symmetry, it
supports two distinct topological phases $|\W|\in\{1,0\}$ featuring either a
pair of edge states or none \cite{asboth2016book}. When longer-range hoppings are added, one can either break particle-hole and chiral symmetry if they
connect the same sublattice (even hoppings),  or preserve the symmetries if they
connect different sublattices (odd hoppings). Even hoppings change the
topological class from BDI to AI \cite{tenfoldway}, while odd hoppings do not
change the topological class and allow for larger values of the topological
invariant \cite{longrangeW}. Its maximum value is ultimately fixed by the range
of the hoppings considered $|\W|\leq(R+1)/2$. Finding a closed expression for
$\mathcal{W}$ as a function of the system parameters is in general a hard task
for couplings beyond NNN \cite{phasediag2vec}. In the
appendix~\ref{app:phasediagram}, we show how this can be done for first
and third neighbors. 

\begin{figure}
	\includegraphics[width=\linewidth]{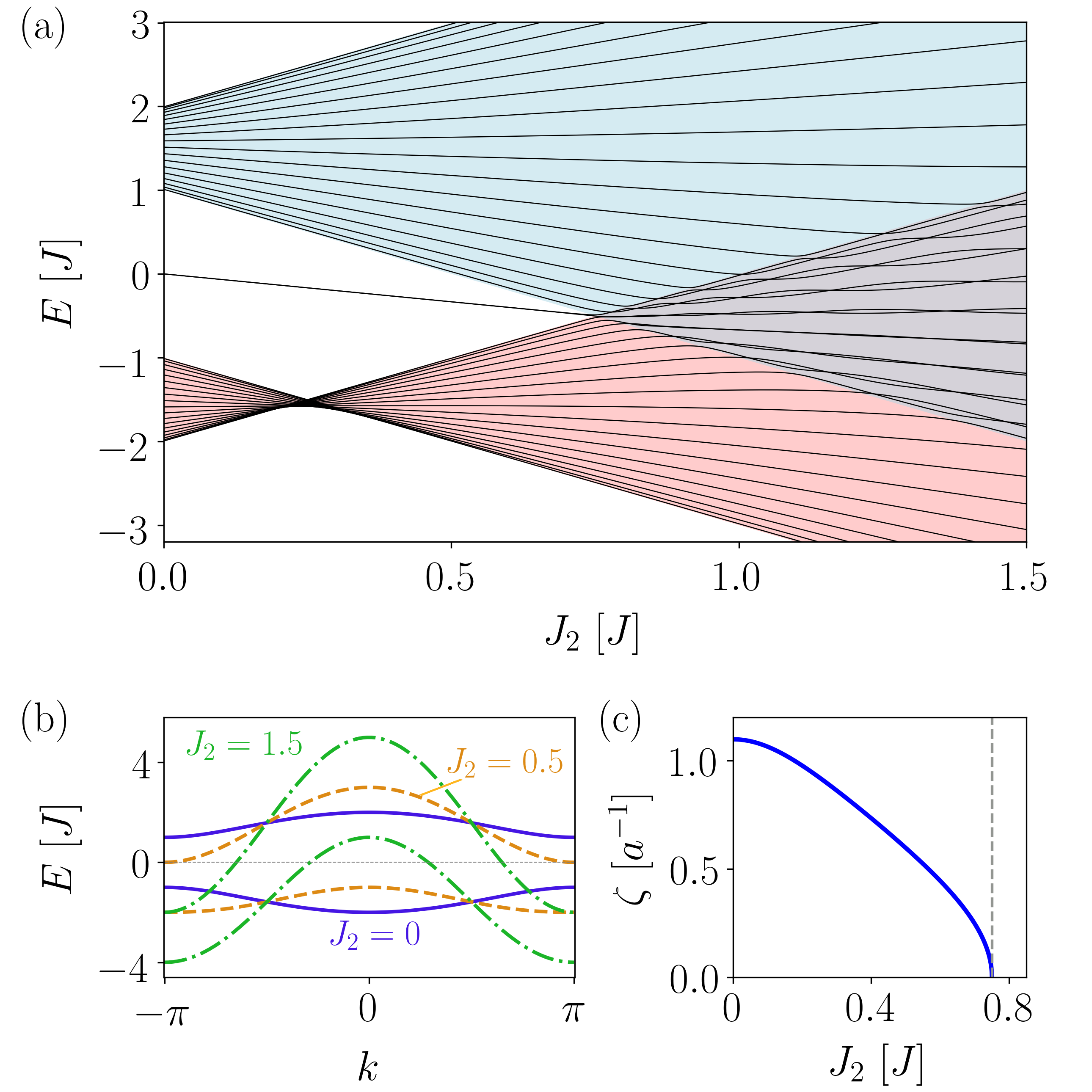}
	\caption{\label{fig:2vec_spectrum_loclength} Spectrum for a chain with up to
  NNN hoppings (a) Spectrum vs. $J_2$ for a finite system with $N=20$ unit cells.
  (b) Spectrum vs. crystal momentum for different values of $J_2$. (c) Inverse
  of localization length vs. $J_2$ for the edge states.  All plots consider a
  chain with $\delta=-0.5$.} 
\end{figure}

With arbitrary long-range hopping amplitudes the system may still feature edge states, but
there is not a one-to-one correspondence between the number of edge states and
the topological invariant: the presence of even hoppings introduces a term
proportional to the identity matrix, which does not change the bulk eigenstates
(leaving $\mathcal{Z}$ unaffected), but modifies the spectrum, making the bands
overlap for sufficiently large values of the hopping amplitudes, changing the
system from insulating to metallic. This is seen in
Fig.~\ref{fig:2vec_spectrum_loclength}(a), where the spectrum, as a function of
the NNN hopping amplitude $J_2$,  makes the edge modes mix with the bulk bands when the
single particle gap closes \cite{metallic2vec}.  

In a finite system, the presence of even hopping amplitudes also affects the spatial
profile of the mid-gap states, as they do not come in chiral pairs anymore.
Their localization length generally increases, diverging when they mix with the
bulk bands. This can be seen already for NNN hoppings: From perturbation theory
one can see that the energy of the edge states varies as ${E_\mathrm{edge}\simeq
-2 J_2 J'_1/J_1}$. Then, looking for solutions of the dispersion relation with
${k=\pi\pm i\zeta}$ we obtain the following expression for the inverse of the
localization length: 
\begin{multline}
  \zeta = \frac{1}{\lambda_\mathrm{loc}} = 
  \mathrm{acosh}\left[\frac{J'_1}{J_1} - \frac{J_1J'_1}{4J^2_2} \right. \\
  \left. + \frac{1}{4J^2_2}\sqrt{4J^2_2(J^2_1 - J'^2_1)+J^2_1J'^2_1}\right] \,,
  \label{eq:loc_length}
\end{multline}
which is plotted in Fig.~\ref{fig:2vec_spectrum_loclength}(c). There, the
divergence of the localization length signals the transition to the metallic
phase.   

In the following, we focus on the effect of disorder, as well as on the
transport properties with NNN present. For convenience, we re-parameterize
$J_1=J(1-\delta)$ and $J'_1=J(1+\delta)$ in terms of the dimerization factor
$\delta$, and the average NN hopping amplitude $J$. In this convention, the topologically
non-trivial (trivial) phase with $|\W|=1$ ($|\W|=0$) occurs for $\delta<0$
($\delta>0$).  

\section{Disorder}

\begin{figure*}[t]
	\includegraphics[width=\linewidth]{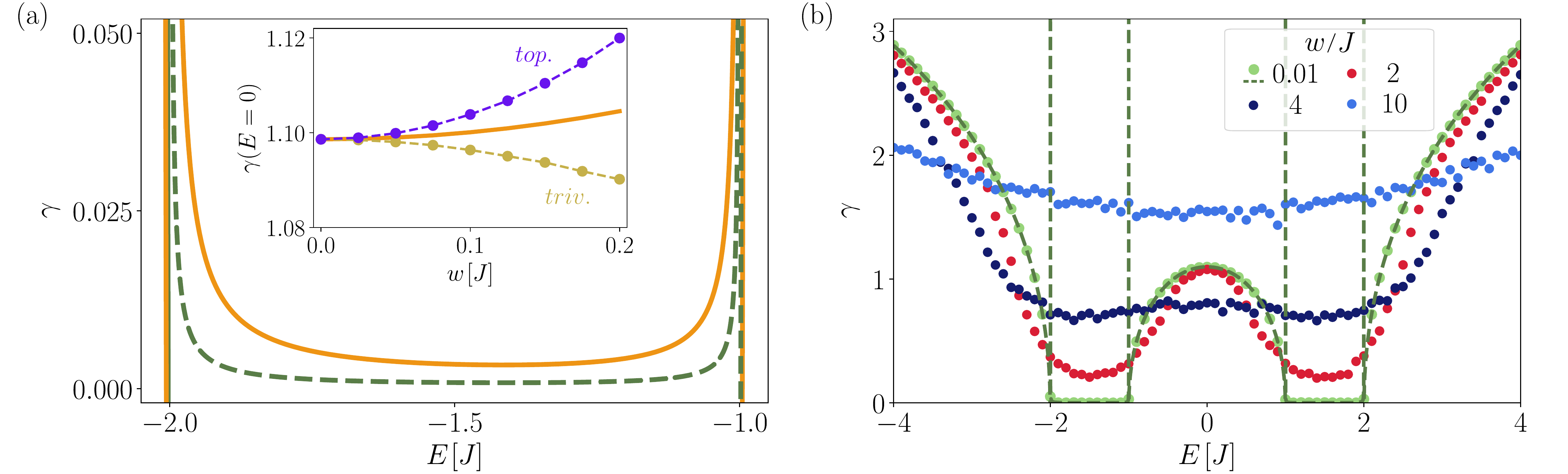}
  \caption{\label{fig:dis_SSH}(a) Comparison between $\gamma_\mathrm{DD}$
  (Eq.~\eqref{eq:Diagonal-Dis-1}, dashed green line) and $\gamma_\mathrm{ODD}$
  (Eq.~\eqref{eq:Off-Diagonal-Dis-1}, orange line) for the states in the lower
  energy band with $\delta = -0.5$ and $w/J = 0.1$. Inset: numerically
  calculated $\gamma_\mathrm{ODD}(E=0)$ in the trivial (yellow dots) and
  topological (purple dots) phase as a function of the disorder strength $w$. 
  The analytic formula Eq.~\eqref{eq:Off-Diagonal-Dis-1} corresponds to the
  continuous orange line. (b) $\gamma_\mathrm{DD}$ vs $E$ for
  different disorder strengths $w$ in the SSH model. The dashed line (dark green)
  corresponds to the analytical approximation (Eq.~\ref{eq:Diagonal-Dis-1}) and
  the dots to the numerical calculation. Parameters: $\delta = -0.5$.} 
\end{figure*}

Diagonal disorder modifies the on-site energies, and therefore acts within the
same sublattice, breaking the chiral symmetry initially present in the standard
SSH model. This modifies the topological phases, leads to a splitting of the
zero-energy modes and produces Anderson localization. On the other hand,
off-diagonal disorder preserves chiral symmetry (well known for the Anderson
model
\cite{soukoulis81,fluctuationFleishman,LogOffdiag1,locdeloc,offdiaglongrange})
and keeps the topological phase well defined for weak disorder. In this case the
bulk electrons also localize, but the edge states remain gapless until disorder
is of the order of the dimerization factor $w \sim \delta$. 

To gain insight into the localization properties in the standard SSH model, we
have calculated a moment expansion of the Lyapunov exponent (LE)
\cite{thesisConductanceRings}: 
\begin{eqnarray}
  \gamma_\mathrm{DD}(E) & \simeq & \log|A|-
  \frac{E^{2}\sigma^{2}}{(E^{2}-4J^{2})(E^{2}-4J^{2}\delta^{2})}
  \label{eq:Diagonal-Dis-1}\\
  \gamma_\mathrm{ODD}(E) & \simeq & \log|A|-
  \frac{h(E)\sigma^{2}}{2(A^{2}-1)^{2}J_1 ^2 J_1^{\prime 2}}
  \label{eq:Off-Diagonal-Dis-1}
\end{eqnarray}
where
\begin{align}
  A(E)&=\left[f(E)\pm\sqrt{f(E)^2-4J_1^2{J'}_1^2}\right]/2J_1J'_1 \,, \\
  f(E)&=E^2-J_1^2-{J'}_1^2 \,, \\
  \begin{split}
  h(E)&=(A^4-8A^2-1)J_1^2-4A(A^2+3)J_1J_1^\prime \\
    &\qquad+(A^4-8A^2-1){J'}_1^2 \,.
  \end{split}
\end{align}
(Details of the calculation can be found in the Appendix~\ref{app:lyapunov}).
Eq.\eqref{eq:Diagonal-Dis-1} corresponds to the case of weak diagonal disorder,
and Eq. \eqref{eq:Off-Diagonal-Dis-1} corresponds to the case of weak
off-diagonal disorder. Fig.~\ref{fig:dis_SSH}(a) shows a comparison between the
two analytic formulas. They display similar behavior, with the exception that
ODD produces stronger localization. This is expected because values of the
hopping amplitudes close to zero lead to full localization, and therefore disorder in the
hopping amplitudes should produce stronger effects, for the same disorder
strength. 

Eqs. \eqref{eq:Diagonal-Dis-1} and \eqref{eq:Off-Diagonal-Dis-1} recover the
expressions for the Anderson model obtained by the same method:
$\gamma(E)\propto -\sigma^2/(E^2-4 J^2)$ in the limit $\delta\rightarrow 0$
(valid for both DD and ODD, being the prefactors the only difference between
them). Importantly, as Eqs.\eqref{eq:Diagonal-Dis-1} and
\eqref{eq:Off-Diagonal-Dis-1} do not depend on the sign of $\delta$, the trivial
and topological phases cannot be distinguished by their localization properties
for weak disorder, at least to second order in the perturbative expansion. This
is not surprising because topological properties are not captured by local quantities.

Interestingly, due to the poles structure in Eqs. \eqref{eq:Diagonal-Dis-1} and
\eqref{eq:Off-Diagonal-Dis-1}, there is a crossover region within the gap where
localization is anomalous, changing sign for the states in the gap, which become
less localized for increasing disorder. This is well-known in non-topological
systems, but topological systems provide an extra twist: they display mid-gap
states (only present in the topological phase) that delocalize as disorder
increases, while the bulk states localize more. This is important for
measurements depending only on these mid-gap states, such as edge modes
transport. Nevertheless, it is also important to keep in mind that structural
defects such as impurities and dangling bonds can also produce these localized
states within the gap. However, one could distinguish them with additional
measurements, as for example testing the robustness of the topological states to
remain at zero energy under small perturbations. 

As a check we have calculated numerically the Green's function $G=(E-H)^{-1}$,
and by fitting the disorder-averaged elements $-\log(\langle G_{1,L}\rangle^2)$,
($L=1,\dots,2N$), we have obtained $\gamma(E)$ from its slope.
Fig.~\ref{fig:dis_SSH}(b) shows the LE vs $E$ for different values of DD and a
comparison with the analytical expression for weak disorder. The agreement is
excellent and captures the previously discussed delocalization of midgap states.
For large disorder these phenomenon disappears and all states become more
localized for increasing $w$, as typically observed in Anderson localization.

An interesting property, missing in the perturbative calculation, is shown in
Fig.~\ref{fig:dis_SSH} (inset). It shows that the difference in $\gamma(E=0)$
for the trivial and the topological phase, when ODD is added, is a non-vanishing
function of the disorder strength (with variance always smaller than this
difference). We have checked that this difference remains for different system
lengths ($N=50,100,150$ and $200$), which means that it should remain in the
limit $N\rightarrow \infty$. Furthermore, this difference scales as $\sigma^2$,
which would indicate that it does not correspond to higher orders in the
perturbative expansion. Therefore, these evidences indicate that this
contribution is consequence of the presence of the edge modes. 

\begin{figure}[t]
  \includegraphics[scale=0.4]{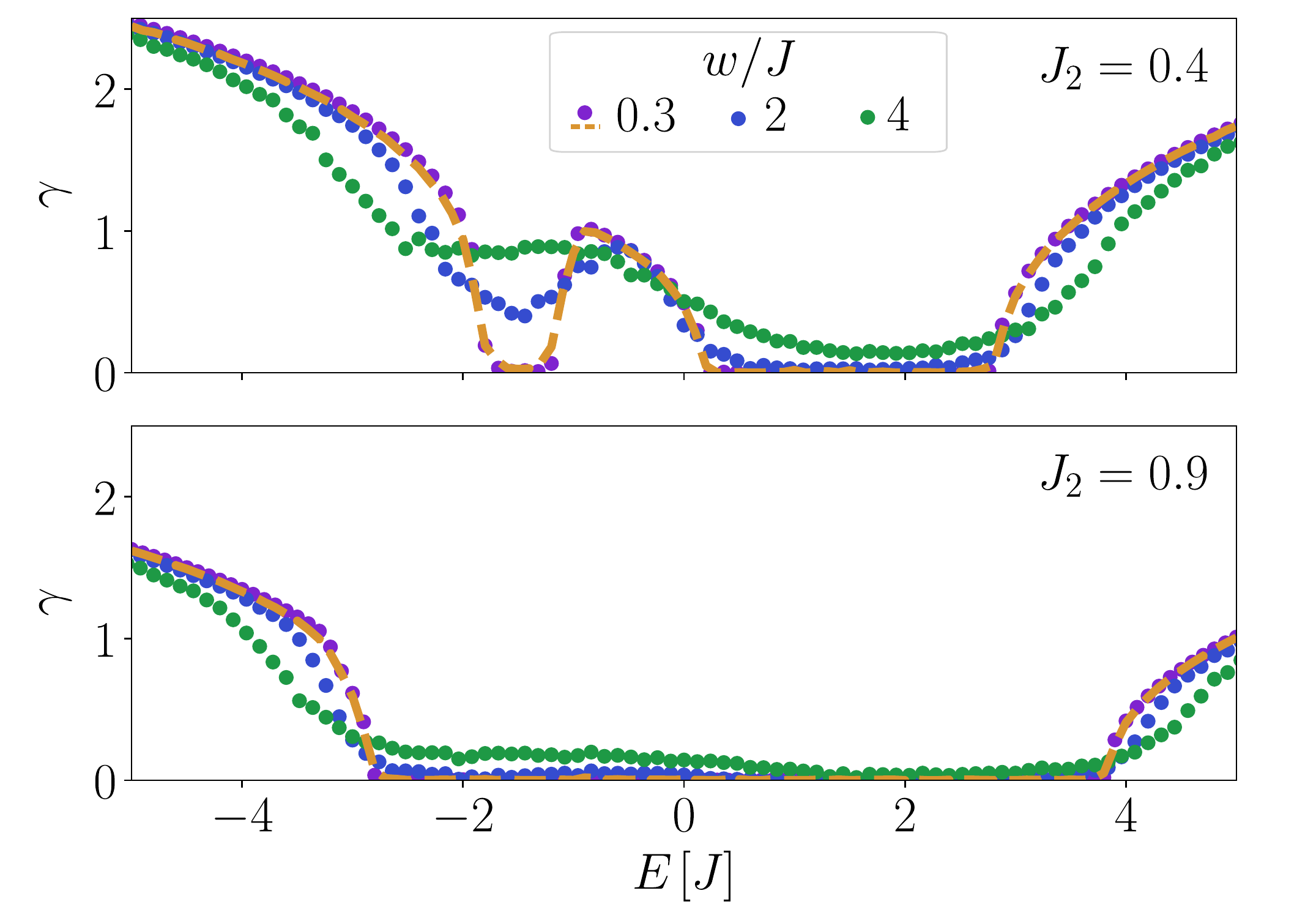}
	\caption{\label{fig:Diagonal-disorder-NNN} Lyapunov exponent vs energy for a
  chain with fixed NN hopping amplitudes and different values of the NNN hopping amplitudes. The
  dots correspond to diagonal disorder, while the dashed orange line
  corresponds to off-diagonal disorder. Parameters: $\delta=-0.5$.}
\end{figure}

When NNN hopping amplitudes are included, the electronic properties can be severely
affected due to the changes in the energy bands, and specially due to the
transition to a metallic phase with electron and hole pockets for large $J_{2}$.
Fig.~\ref{fig:Diagonal-disorder-NNN} shows the value of $\gamma\left(E\right)$
for different values of $J_{2}$ and disorder.  For small $J_{2}$ the chiral
symmetry is weakly broken and the system still displays a gap, see
Fig.~\ref{fig:2vec_spectrum_loclength}. As $J_2$ increases, the bands and the LE
become increasingly asymmetrical. It is interesting to see how DD and ODD
disorder act in a similar way when NNN are present (purple and orange lines in
Fig.~\ref{fig:Diagonal-disorder-NNN}). This is because chiral symmetry is
already broken by $J_2$, and off-diagonal disorder also introduces fluctuations
within the same sublattice \cite{sublattice1}. For large $J_2$ the system
becomes metallic and although the LE behaves similarly for DD and ODD, we
demonstrate that transport in the trivial and in the topological phase can be
quite different.    

\section{Single-particle transport}

Coupling the ends of the SSH chain to voltage-biased leads allows for particle 
transport. We use second-order perturbation theory to integrate out 
the leads, and obtain a master equation for the reduced density matrix of the 
chain $\rho$, which in the infinite bias regime 
assumes the following Lindblad form \cite{breuer2007book, benito2016}:
\begin{equation}
    \dot{\rho}=\mathcal{L}\rho \equiv 
    -i[H_R, \rho] + \Gamma_L\mathcal{D}(c^\dagger_1)\rho + 
    \Gamma_R\mathcal{D}(c_{2N})\rho \,,
    \label{eq:mastereq}
\end{equation}
where $\mathcal{D}(A)\rho \equiv A\rho A^\dagger - \{A^\dagger A, \rho\}/2$. 
The last two terms in the r.h.s of Eq. \eqref{eq:mastereq} correspond to 
incoherent tunneling of particles from the left lead into the first site of the 
chain at rate $\Gamma_L$, and tunneling out from the last site of the chain to
the right lead at rate $\Gamma_R$. 

We analyze transport considering there is at most one particle in the 
system. This is the case if the interaction between particles inside the chain
is strong enough such that higher chain occupancies are forbidden. The current
in the stationary regime can be computed as $I=\tr{\mathcal{J}\rho_0}$, where
$\rho_0$ is the stationary solution of the master equation \eqref{eq:mastereq}
and $\mathcal{J}\rho = \Gamma_L c^\dagger_1\rho c_1$ is the current 
superoperator. 

\begin{figure}
    \centering
    \includegraphics[width=\linewidth]{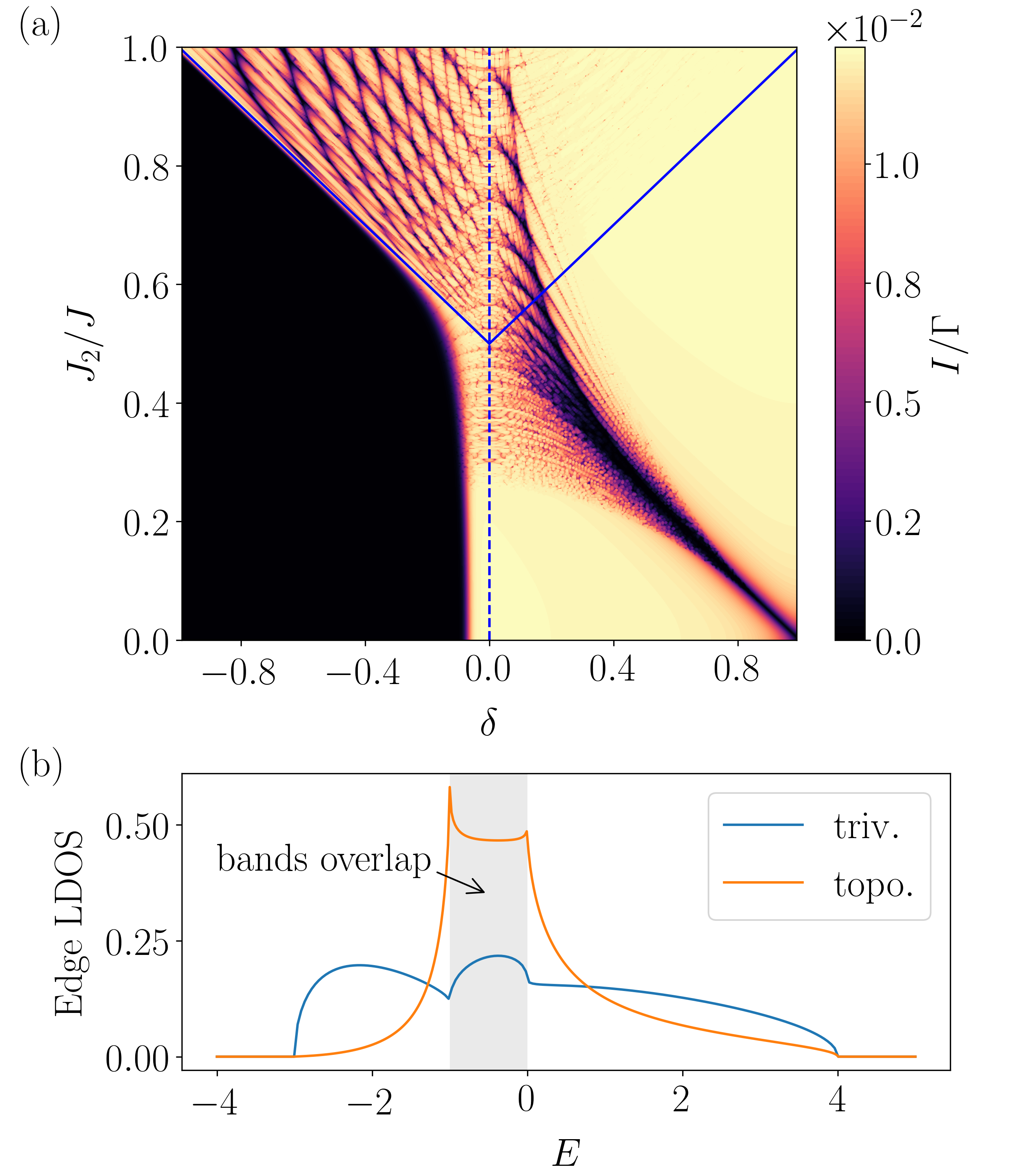}
    \caption{(a) Current with $\Gamma_L=\Gamma_R=0.1J$, for a chain with $N=40$
    dimers. The dashed vertical line marks the topological phase transition, 
    while the upper triangle delimited by continuous lines marks the region 
    where the system is metallic. (b) Local density of states at the ending 
    site for a semi-infinite system with $J_2=J$ and $\delta=-0.5$ (topological 
    case) or $\delta=0.5$ (trivial case).} 
    \label{fig:current_and_LDOS}
\end{figure}

In Fig.~\ref{fig:current_and_LDOS}, we show the current in a pristine dimer
chain with hopping amplitudes up to 2nd neighbors as a function of the different hopping
amplitudes. In the insulating regime in the topologically non-trivial phase
($\delta<0$), we find the \textit{topological edge-state blockade} already
studied in Refs.~\onlinecite{benito2016,Ruocco2017}. As $J_2$ increases, the
blockade remains up to the transition to the metallic phase. In the trivial
region ($\delta>0$) the current vanishes along the line $J_2\simeq -\delta J/2 +
J/2$, which corresponds to a configuration where the lower energy band is almost
flat. For large enough $J_2$ the system is in the metallic phase and the current
shows a pattern of dips in both the trivial and topological phases produced by
exact crossings of energy levels with opposite parity (see the spectrum shown in
Fig.~\ref{fig:2vec_spectrum_loclength}).  Whenever two states with opposite
parity become degenerate, a superposition of both with zero occupation at the
ending site of the chain becomes a steady state of the system.  A particle in
this superposition prevents any new particle from tunneling into the chain, and
it cannot escape to the drain, thus blocking the current.  Remarkably, most of
these degeneracies occur for the same values of $|\delta|$ and $J_2/J$ in the
topological and trivial regimes. Nonetheless, most of the dips in the trivial
region cannot be appreciated in Fig.~\ref{fig:current_and_LDOS} since they are
much sharper than those in the topological regime (see
Fig.~\ref{fig:minimalmodel} in the appendix \ref{app:currentanalysis}). The
reason why resides in the way these states with opposite parity split near the
degeneracy, and also in the relative weight they have on the ending sites of the
chain as we demonstrate in appendix \ref{app:currentanalysis} by analyzing a
minimal model with three energy levels.  This latter fact relates the local
density of states at the ending sites of the chain \cite{moller2012}, shown in
Fig.~\ref{fig:current_and_LDOS}(b), with the features observed in the current. 

We have also computed the Fano factor, which is a measure of the shot noise in
the transport process. Previous studies showed that for the standard SSH model
the Fano factor is approximately equal to 1 in the topological regime
\cite{benito2016}, meaning that transport is purely Poissonian. We find that for
finite $J_2$ it is generally larger than 1 and presents clear differences
between the topological and trivial phases. In the metallic region, it presents
a characteristic peak-inside-a-dip shape at the exact crossings discussed above, 
that can again be related to the edge-LDOS, as we show in
appendix~\ref{app:currentanalysis}.

\begin{figure}
    \centering
    \includegraphics[width=\linewidth]{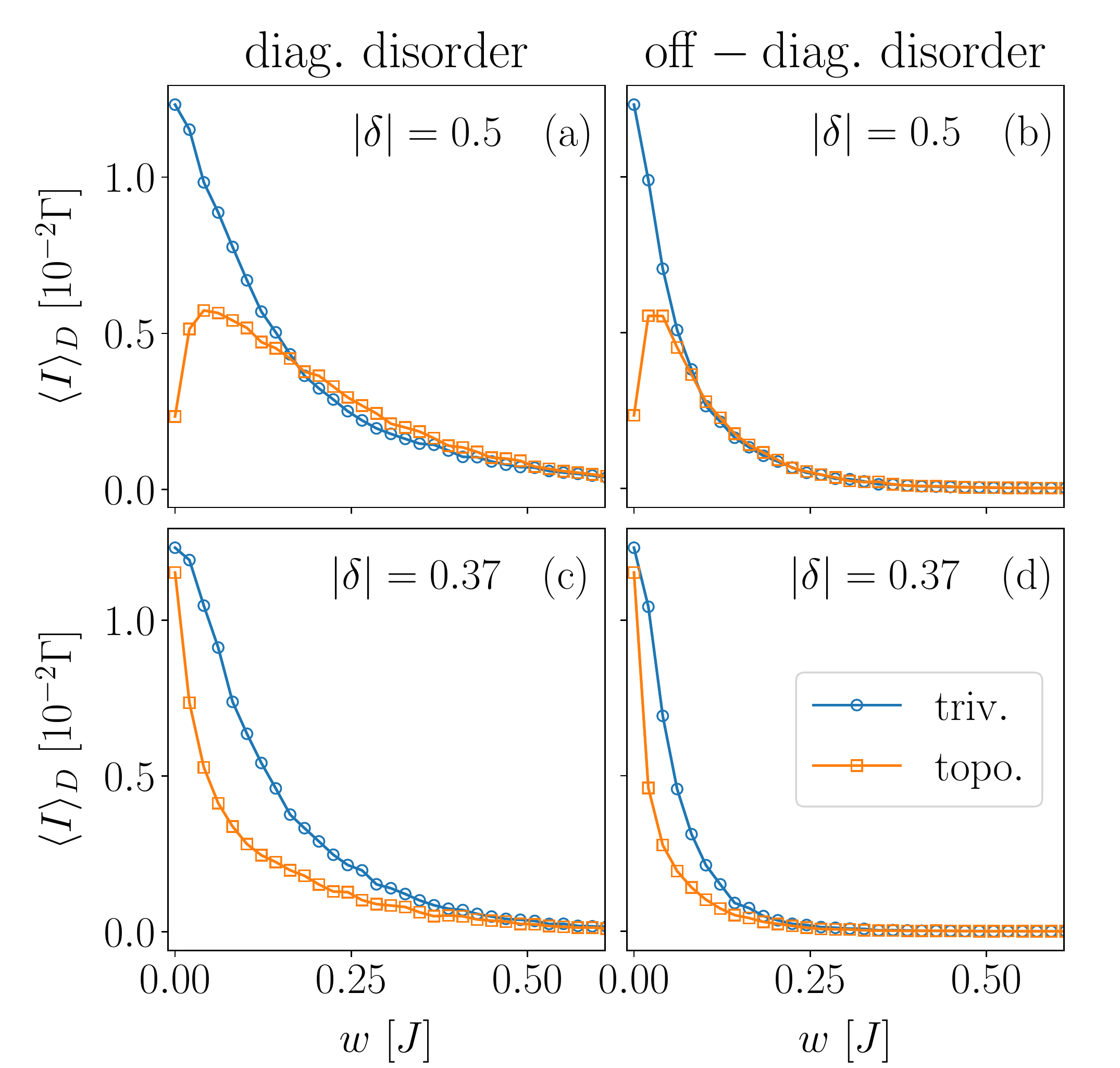}
    \caption{Current as a function of the disorder strength for $J_2=0.9J$,
    $N=40$, and $\Gamma_L=\Gamma_R=0.1J$. Both diagonal (a, c) and off-diagonal
    disorder (b, d) is considered. The distribution of $\epsilon_{ij}$ is the
    same as the one used in the other figures. Each point has been obtained by
    averaging over a total of $10^3$ instances of disorder. The parameters have
    been chosen such that on the upper plots, the current without disorder has a
    dip in the topological regime, while on the lower plots, the current shows a
    maximum in the topological regime.}
    \label{fig:current_vs_disorder}
\end{figure}

The effect of disorder on the current, when hoppings up to NNN are included, is
similar for both DD and ODD, see Fig.~\ref{fig:current_vs_disorder}. In the
topological phase, near the exact crossings, the current shows a non-monotonic
behavior, it increases for small disorder and decreases for large disorder. In
the trivial phase, by contrast, the current decays monotonically with increasing
disorder.  This can be understood as follows: For zero disorder, near the exact
crossings, the current is much smaller in the topological case than in the
trivial, as it is carried mainly by almost-degenerate states with opposite
parity, while in the trivial case it is carried equally by all states in the
spectrum. Therefore, for small disorder, the current in the topological phase
increases as disorder lifts the degeneracies that are blocking the current. On
the other hand, in the trivial case the current decreases due to the increasing
localization of the states. For large disorder the current decreases in both
phases as expected.   

\section{Conclusions}
We have studied a generalized SSH model including long-range hopping amplitudes and
disorder.  We have shown that the effect of hopping amplitudes connecting sites within the
same sublattice, and those connecting sites of different sublattices is very
different. The reason is that the former breaks particle-hole and chiral
symmetry, changing the topological class, while the later maintains chiral
symmetry and allows to increase the value of the winding number. With both types
of hopping processes, space inversion symmetry forces the topological invariant to have
quantized values, but the bulk-edge correspondence breaks-down. This is clear
when NNN hopping amplitudes dominate, producing a metallic phase and the merging of the
edge states with the bulk bands. 

We have investigated the role of disorder using the Lyapunov exponent. It shows
that for NN hoppings only, DD and ODD localize the bulk electrons, but their
effect on mid-gap states is different, with a crossover from reduced to
increased localization as a function of disorder strength. In addition, our
numerical calculations find an extra contribution to $\gamma(0)$, as a function
of ODD strength, that leads to a difference between the trivial and topological
phase. This contribution is linked to the presence of edge modes.  When
NNN hoppings are added, chiral symmetry is broken and fluctuations in $J_2$ act similarly
to DD. Nevertheless we have shown that transport measurements for weak disorder
can still distinguish between phases with different $\mathcal{W}$. Furthermore,
the current in the metallic phase shows interesting features that also allow to
differentiate between them.  

Our findings could be observed experimentally with arrays of quantum dots
\cite{zajac2016}, in which the large Coulomb repulsion needed to keep at most
one electron in the system could be engineered by capacitively coupling all
dots together. In quantum dots, first-neighbor hoppings are of the order of $10\hspace{2pt}meV$, while disorder strongly depends on the material and sample configuration. 

Also, our results are relevant to the transport of excitations 
in analogue systems, which could be implemented in platforms such as
trapped ions \cite{nevado2017} or cold atoms 
\cite{sshoffdiagexp, disordercoldatoms}.  

\section*{Acknowledgements}
This work was supported by the Spanish Ministry of Economy and Competitiveness 
through Grant No.MAT2014-58241-P and Grant MAT2017-86717-P. M. Bello acknowledges the FPI 
program BES-2015-071573, \'A. G\'omez-Le\'on acknowledges the Juan de la 
Cierva program and Beatriz P\'erez-Gonz\'alez acknowledges the FPU program FPU17/05297.

\appendix
\section{Phase diagram with long-range hopping amplitudes} \label{app:phasediagram}
We derive here the phase diagram for a chain with up to fourth-neighbor
couplings ($R\leq 4$). It only depends on the odd hopping amplitudes, i.e., on
the ratios $x=J_1/J'_1$, $y=J'_3/J'_1$ and $z=J_3/J'_1$. The two-dimensional
surfaces that delimit the regions with definite winding number are those points
in parameter space for which the system of equations $d_x(k)=d_y(k)=0$ has
solution for some $k\in[-\pi,\pi]$. These turn out to be the equations of two
planes $P_1(x,y,z)=0$, $P_2(x,y,z)=0$, and a quadric 
$Q(x,y,z)=0$, with
\begin{gather}
  P_1(x,y,z)=1+x+y+z\,,\\
  P_2(x,y,z)=1-x-y+z\,,\\
  Q(x,y,z)=z-z^2-xy+y^2\,.
\end{gather}
They divide the parameter space in finitely many regions. Computing the winding
number inside each of these regions, we obtain the whole phase diagram. For
positive hopping amplitudes the results are summarized in the following table 
(note that the plane $P_1$ does not appear).
\begin{center}
  \begin{tabular}{|C{0.8cm}|C{5.5cm}|}
    \hline
    $|\W|$ & Conditions \\ 
    \hline\hline
    0 & ($Q>0$ and $P_2>0$) or \newline ($Q<0$, $P_2>0$ and $|y-x|>2|z|$) \\
    \hline
    1 & $P_2<1$ \\
    \hline
    2 & $Q<0$, $P_2>0$ and $|y-x|\leq 2|z|$ \\
    \hline
  \end{tabular}
\end{center}

\section{Calculation of the Lyapunov exponent} \label{app:lyapunov}

In the weak disorder limit, the analytic expressions for the Lyapunov exponent
can be obtained perturbatively from the equation of motion (EOM). This can be
done by introducing the parameter $\lambda\ll 1$ in the disorder term as $ E -
\lambda \epsilon^{\alpha}_m $ (where $\alpha$ is the sublattice index, $A$ and
$B$), and $J_{ij}+\lambda \epsilon_{ij}$, for the diagonal and off-diagonal case
respectively. As the SSH model is bipartite, each sublattice has an associated
EOM, but they can be combined into a single one (for instance, for sublattice
A). 

By defining $R_m = a_m/a_{m-1}$, where $a_m$ is the probability amplitude of the
electronic wave function in the $m^{\mathrm{th}}$ A-site of the chain, the
Lyapunov exponent can be written as 
\begin{equation}
\gamma(E) = \lim_{M \rightarrow \infty}\frac{1}{M}
  \left\langle \sum_{m=1}^{M}\ln|R_m| \right\rangle \,,
\end{equation}
where is $M$ is the number of cells in the chain and $\langle \cdot \rangle$
denotes configuration average. To find the Lyapunov exponent, $R_m$ can be
expanded over the parameter $\lambda$ in the following way 
\begin{equation}
	R_m = A e^{\lambda B_m+\lambda^2 C_m + \lambda^3 D_m+...}\,,
\end{equation}
which in the limit $\lambda\ll 1$ yields
\begin{equation}
\gamma(E) = \lim_{M \rightarrow \infty}\frac{1}{M}\sum_{m=1}^M \bigg[ \langle \ln|A|\rangle + \lambda \langle B_m \rangle + \lambda^2 \langle C_n \rangle + ... \bigg]\,.
\end{equation}

By obtaining the leading orders of $\lambda$ from the corresponding EOM, one can
calculate $\langle \ln|A|\rangle$, $\langle B_m \rangle$, and $\langle C_n
\rangle$.  The $\lambda^0$ order of the combined EOM let us solve $A$ as a
function of the hopping amplitudes and energy, which yields the following result
for both the DD and ODD case,   
\begin{equation}
A_{\pm} = \frac{E^2-J_1^2-J_1^{'2}\pm \sqrt{(E^2-J_1^2-J_1^{'2})^2-4J_1^2 J_1^{'2}}}{2J_1 J_1'} \,.
\end{equation}
where $A_{-}$ is valid within the band and the gap energy regions and $A_{+}$
outside the bands. This choice is necessary to ensure $\gamma(E)$ is positive
for all $E$. Thus, $\gamma(E)=\lim_{M \rightarrow \infty}\frac{1}{M}\sum_{m=1}^M
\langle \ln|A|\rangle$ defines the localization properties of the pristine SSH
model. As expected, $\gamma(E)$ is zero within the energy bands and non-zero
outside. Particularly, $\gamma(0) = 1/l = \log{|J_1/J_1'|}$ in the
topological phase, which matches the well-known result for the localization
length of the edge states \cite{delplace2011}.  

Next, $\langle B_n \rangle$ can be obtained through the $\lambda^1$-equation
upon averaging. For DD, we choose $\langle \epsilon_m \rangle = 0$ for both
sublattices, as the chemical potential in the pristine system has been set to
zero for all sites and the randomness induced by disorder is assumed to
fluctuate around that value. For ODD, we set $\langle\epsilon_{ij}\rangle = 0$.
In both cases, this results in a recursive equation for $\langle B_m \rangle $
which can be solved with $\langle B_m\rangle = 0$ without loss of generality. 

Then, the $\langle C_m \rangle$ is calculated from the $\lambda^2$-equation
after averaging. At this point, it is important to note that disorder, as
introduced in the system, displays no correlations, and hence $\langle
\epsilon^\alpha_m \epsilon^\beta_n \rangle = \delta_{\alpha
\beta}\delta_{mn}\sigma^2$ for DD and $\langle \epsilon_{ij}\epsilon_{mn}\rangle
= \delta_{ij}\delta_{mn}\sigma^2$ for ODD, where $\sigma^2$ is the standard
deviation of the probability distribution followed by the random terms.  

Finally, the Lyapunov exponent can be written as $\gamma(E)=\ln{|A|}+\langle
C_m\rangle$ for both cases. For DD, this yields 
\begin{equation}
\gamma(A) = \ln{|A|} - \frac{A(A J_1 + J_1')(J_1 + A J_1')}{(A^2 - 1)^2
  J_1^2}\sigma^2 \,.
\end{equation}

For ODD, one obtains
\begin{multline}
	\gamma(A) = \ln|A| + \bigg[ \frac{(A^4 - 8 A^2 -1)J_1^2- 4 A(A^2+3)J_1 J_1' }{2(A^2-1)^2 J_1^2 J_1^{'2}}\\
	 + \frac{(A^4-8A^2-1)J_1^{'2}}{2(A^2-1)^2 J_1^2 J_1^{'2}}\bigg] \sigma^2\,.
\end{multline}

As a function of the energy, it is easy to compare the Lyapunov exponent for ODD and DD case, 
\begin{align}
	\gamma_\mathrm{DD}(E) &= \ln{|A|} - \kappa(E)\sigma^2\,,\\
	\gamma_\mathrm{ODD}(E) &= \ln{|A|} - \kappa(E)\sigma^2 + \xi(E)\sigma^2\,,
\end{align}
where
\begin{equation}
	\kappa(E)=\frac{4 E^2}{(4 J^2-E^2)(4 J^2 \delta^2- E^2)}\,,
\end{equation}
and
\begin{equation}
	\xi(E)=\frac{\big[8J^2 \delta^2 - E^2(1+\delta^2)\big]\sqrt{(4J^2-E^2)(4J^2 \delta^2-E^2)}}{J^2 (4 J^2-E^2)(4 J^2 \delta^2- E^2)(\delta^2-1)^2} \,.
\end{equation}
The term $\xi(E)$ does not contribute in the band region, since the radicand is
negative and hence the square root factor is imaginary. 

\section{Analysis of the current and Fano factor} \label{app:currentanalysis}
Explaining all the features the current and the Fano factor display is a hard
task that requires knowledge of the full eigenbasis of the system. Here we
attempt to give a qualitative understanding of the observed behavior near the
exact crossings of the energy spectrum, which can be achieved considering a
minimal single-particle model comprised of two almost-degenerate states with
opposite parity $\ket{\pm}$ and a third level $\ket{C}$ which is highly detuned
from the other two, such that it can be regarded as an exclusive channel for
transport. 

\begin{figure}[t]
  \centering
  \includegraphics[width=\linewidth]{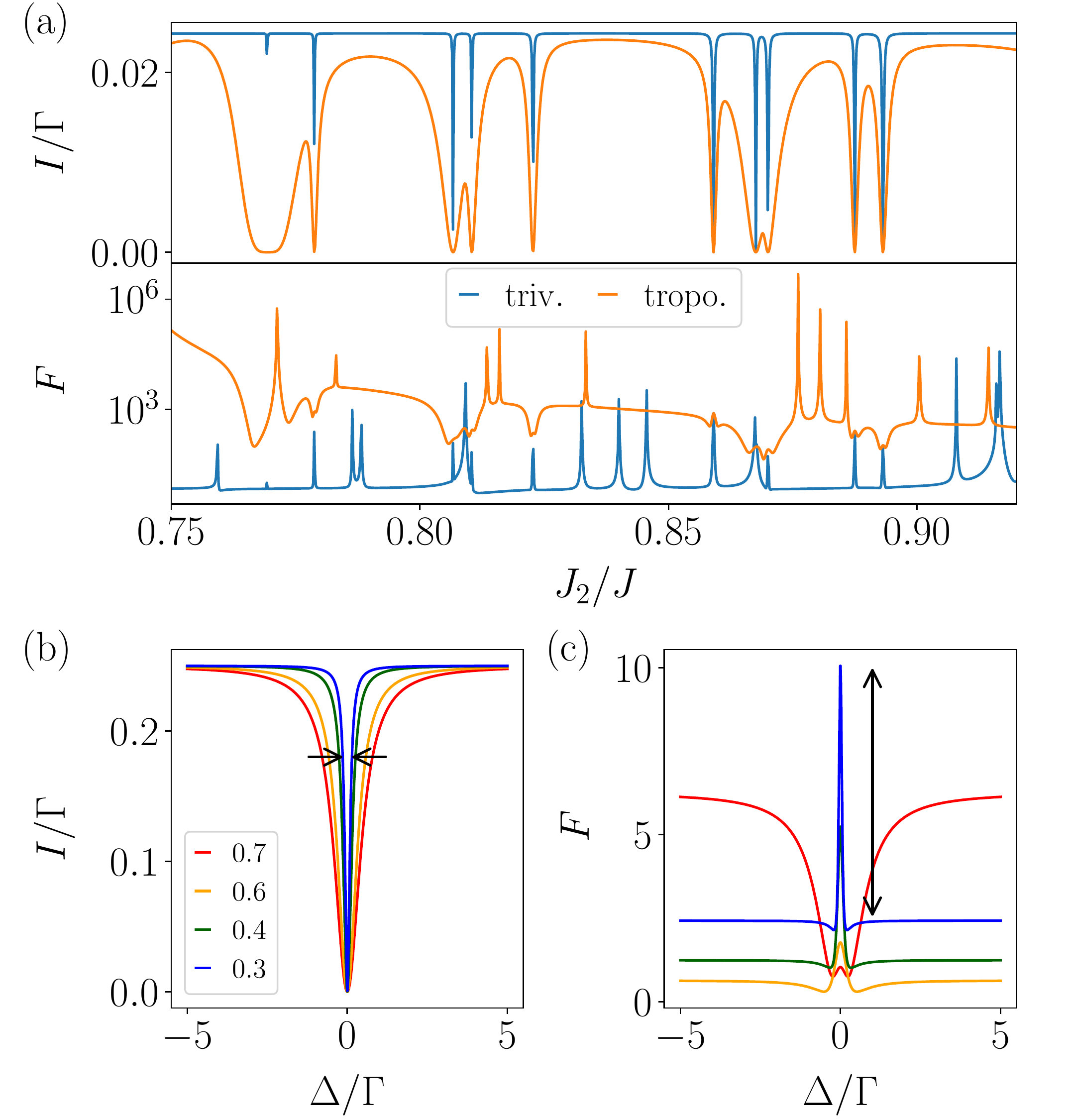}
  \caption{(a) Current and Fano factor as a function of the NNN hopping amplitude
  for $\delta=\pm0.5$, in the metallic region. The rest of the parameters are
  the same as in Fig.~\ref{fig:current_and_LDOS}. (b-c) Plot of the current and
  Fano factor in the minimal model for different values of $\alpha=\beta$ with
  $\Gamma_R=\Gamma_L\equiv\Gamma$. In (b) the width of the dip decreases as
  $\alpha^2$ and $\beta^2$ become smaller (see black arrow). In (c) the height
  of the peak (black arrow) increases as $\alpha^2$ and $\beta^2$ become
  smaller.}
  \label{fig:minimalmodel}
\end{figure}

We can solve analytically the master equation for this small system noting that:

\begin{itemize}
  \item The coherences between states with different number of particles evolve
    independently of the coherences between states with the same number of
    particles and the level populations.
  \item We can do a rotating-wave approximation in which we neglect coherences
    between eigenstates that are far off-resonance, that is, when their energy
    difference $(\epsilon_\mu-\epsilon_\nu)\gg \Gamma_{L,R}$.
\end{itemize}

Thus, in the particular model under consideration, to a good approximation we
can include only the coherences between the $\ket{\pm}$ states. Appart from the
splitting between these two states $\Delta$, the other parameters that enter the
master equation is the weight of the states at the ending sites of the chain:
\begin{gather}
  \bra{0}c_1\ket{+}=\bra{0}c_{2M}\ket{+}=\alpha\,, \\
  \bra{0}c_1\ket{-}=-\bra{0}c_{2M}\ket{-}=\beta\,, \\
  \left|\bra{0}c_1\ket{C}\right|=
  \left|\bra{0}c_{2M}\ket{C}\right|=\sqrt{1-\alpha^2-\beta^2}\,.
\end{gather}
The parity of state $\ket{C}$ is irrelevant regarding the final result. Solving
the master equation in the basis
$\{\ket{0}\bra{0},\ket{+}\bra{+},\ket{-}\bra{-},\ket{+}\bra{-},\ket{-}\bra{+},\ket{C}\bra{C}\}$,
($\ket{0}$ denotes the vacuum state of the system) we obtain for the current 
\begin{equation}
  I = \frac{\Delta^2\Gamma_L\Gamma_R}
  {\Delta^2(3\Gamma_L+\Gamma_R)+(\alpha^2+\beta^2)^2\Gamma_L\Gamma^2_R}\,.
\end{equation}
It goes from a finite value to zero when the states $\ket{\pm}$ become
degenerate ($\Delta=0$). Furthermore, the width of the dip is proportional to
$(\alpha^2+\beta^2)$. For the Fano factor, we will just give formulas for its
value at $\Delta=0$, and in the limit $\Delta\gg\Gamma_{L,R}$, 
\begin{align}
  \lim_{\Delta/\Gamma_{L,R}\to 0} F & =
  \frac{1}{2}\left(\frac{1}{\alpha^2}+\frac{1}{\beta^2}-2\right)\,, \\
  \begin{split}
    \lim_{\Delta/\Gamma_{L,R}\to \pm\infty} F & = 
    1 - \frac{6\Gamma_L}{3\Gamma_L+\Gamma_R}+
    \\ \frac{2\Gamma_L\Gamma_R}{(3\Gamma_L+\Gamma_R)^2} &
    \left(\frac{1}{\alpha^2}+\frac{1}{\beta^2}+\frac{1}{1-\alpha^2-\beta^2}
    \right) \,.
  \end{split}
\end{align}
They come in handy when analyzing the results shown in
Fig.~\ref{fig:minimalmodel}. First, the value of the Fano factor far from the
resonances ($\Delta\gg\Gamma_{L,R}$) is minimal for the case
$1/\alpha^2=1/\beta^2=1/3$, i.e., when all the single-particle states have equal
weigth on the ending sites of the chain. This explains why on average the Fano
factor is larger in the topological phase than in the trivial, as the edge LDOS
is much more regular in this latter case. Furthermore, the minimal model
predicts the appearance of a peak in the Fano factor at the exact crossings,
whose height relative to the base is given by
$\left[-4+3/\alpha^2+3/\beta^2-1/(1-\alpha^2-\beta^2)\right]/8$ in the case $\Gamma_{L}=\Gamma_{R}$.  It increases
as the edge occupation of the states $\ket{\pm}$ decreases, which is also in
accordance with the computed edge LDOS.

\bibliography{longrangeSSH}

\begin{thebibliography}{38}%
\makeatletter
\providecommand \@ifxundefined [1]{%
 \@ifx{#1\undefined}
}%
\providecommand \@ifnum [1]{%
 \ifnum #1\expandafter \@firstoftwo
 \else \expandafter \@secondoftwo
 \fi
}%
\providecommand \@ifx [1]{%
 \ifx #1\expandafter \@firstoftwo
 \else \expandafter \@secondoftwo
 \fi
}%
\providecommand \natexlab [1]{#1}%
\providecommand \enquote  [1]{``#1''}%
\providecommand \bibnamefont  [1]{#1}%
\providecommand \bibfnamefont [1]{#1}%
\providecommand \citenamefont [1]{#1}%
\providecommand \href@noop [0]{\@secondoftwo}%
\providecommand \href [0]{\begingroup \@sanitize@url \@href}%
\providecommand \@href[1]{\@@startlink{#1}\@@href}%
\providecommand \@@href[1]{\endgroup#1\@@endlink}%
\providecommand \@sanitize@url [0]{\catcode `\\12\catcode `\$12\catcode
  `\&12\catcode `\#12\catcode `\^12\catcode `\_12\catcode `\%12\relax}%
\providecommand \@@startlink[1]{}%
\providecommand \@@endlink[0]{}%
\providecommand \url  [0]{\begingroup\@sanitize@url \@url }%
\providecommand \@url [1]{\endgroup\@href {#1}{\urlprefix }}%
\providecommand \urlprefix  [0]{URL }%
\providecommand \Eprint [0]{\href }%
\providecommand \doibase [0]{http://dx.doi.org/}%
\providecommand \selectlanguage [0]{\@gobble}%
\providecommand \bibinfo  [0]{\@secondoftwo}%
\providecommand \bibfield  [0]{\@secondoftwo}%
\providecommand \translation [1]{[#1]}%
\providecommand \BibitemOpen [0]{}%
\providecommand \bibitemStop [0]{}%
\providecommand \bibitemNoStop [0]{.\EOS\space}%
\providecommand \EOS [0]{\spacefactor3000\relax}%
\providecommand \BibitemShut  [1]{\csname bibitem#1\endcsname}%
\let\auto@bib@innerbib\@empty
\bibitem [{\citenamefont {Peierls}(1955)}]{Peierls1955}%
  \BibitemOpen
  \bibfield  {author} {\bibinfo {author} {\bibfnamefont {R.}~\bibnamefont
  {Peierls}},\ }\href@noop {} {\emph {\bibinfo {title} {Quantum theory of
  solids}}}\ (\bibinfo  {publisher} {Oxford University Press},\ \bibinfo {year}
  {1955})\BibitemShut {NoStop}%
\bibitem [{\citenamefont {Jackiw}\ and\ \citenamefont
  {Rebbi}(1976)}]{Jackiv1976}%
  \BibitemOpen
  \bibfield  {author} {\bibinfo {author} {\bibfnamefont {R.}~\bibnamefont
  {Jackiw}}\ and\ \bibinfo {author} {\bibfnamefont {C.}~\bibnamefont {Rebbi}},\
  }\href {\doibase 10.1103/PhysRevD.13.3398} {\bibfield  {journal} {\bibinfo
  {journal} {Phys. Rev. D}\ }\textbf {\bibinfo {volume} {13}},\ \bibinfo
  {pages} {3398} (\bibinfo {year} {1976})}\BibitemShut {NoStop}%
\bibitem [{\citenamefont {Su}\ \emph {et~al.}(1979)\citenamefont {Su},
  \citenamefont {Schrieffer},\ and\ \citenamefont {Heeger}}]{SSH1979}%
  \BibitemOpen
  \bibfield  {author} {\bibinfo {author} {\bibfnamefont {W.~P.}\ \bibnamefont
  {Su}}, \bibinfo {author} {\bibfnamefont {J.~R.}\ \bibnamefont {Schrieffer}},
  \ and\ \bibinfo {author} {\bibfnamefont {A.~J.}\ \bibnamefont {Heeger}},\
  }\href {\doibase 10.1103/PhysRevLett.42.1698} {\bibfield  {journal} {\bibinfo
   {journal} {Phys. Rev. Lett.}\ }\textbf {\bibinfo {volume} {42}},\ \bibinfo
  {pages} {1698} (\bibinfo {year} {1979})}\BibitemShut {NoStop}%
\bibitem [{\citenamefont {Su}\ \emph {et~al.}(1980)\citenamefont {Su},
  \citenamefont {Schrieffer},\ and\ \citenamefont {Heeger}}]{ssh1980}%
  \BibitemOpen
  \bibfield  {author} {\bibinfo {author} {\bibfnamefont {W.~P.}\ \bibnamefont
  {Su}}, \bibinfo {author} {\bibfnamefont {J.~R.}\ \bibnamefont {Schrieffer}},
  \ and\ \bibinfo {author} {\bibfnamefont {A.~J.}\ \bibnamefont {Heeger}},\
  }\href {\doibase 10.1103/PhysRevB.22.2099} {\bibfield  {journal} {\bibinfo
  {journal} {Phys. Rev. B}\ }\textbf {\bibinfo {volume} {22}},\ \bibinfo
  {pages} {2099} (\bibinfo {year} {1980})}\BibitemShut {NoStop}%
\bibitem [{\citenamefont {Di~Liberto}\ \emph {et~al.}(2014)\citenamefont
  {Di~Liberto}, \citenamefont {Malpetti}, \citenamefont {Japaridze},\ and\
  \citenamefont {Morais~Smith}}]{metallic2vec}%
  \BibitemOpen
  \bibfield  {author} {\bibinfo {author} {\bibfnamefont {M.}~\bibnamefont
  {Di~Liberto}}, \bibinfo {author} {\bibfnamefont {D.}~\bibnamefont
  {Malpetti}}, \bibinfo {author} {\bibfnamefont {G.~I.}\ \bibnamefont
  {Japaridze}}, \ and\ \bibinfo {author} {\bibfnamefont {C.}~\bibnamefont
  {Morais~Smith}},\ }\href {\doibase 10.1103/PhysRevA.90.023634} {\bibfield
  {journal} {\bibinfo  {journal} {Phys. Rev. A}\ }\textbf {\bibinfo {volume}
  {90}},\ \bibinfo {pages} {023634} (\bibinfo {year} {2014})}\BibitemShut
  {NoStop}%
\bibitem [{\citenamefont {Viyuela}\ \emph {et~al.}(2018)\citenamefont
  {Viyuela}, \citenamefont {Fu},\ and\ \citenamefont
  {Martin-Delgado}}]{viyuela2018}%
  \BibitemOpen
  \bibfield  {author} {\bibinfo {author} {\bibfnamefont {O.}~\bibnamefont
  {Viyuela}}, \bibinfo {author} {\bibfnamefont {L.}~\bibnamefont {Fu}}, \ and\
  \bibinfo {author} {\bibfnamefont {M.~A.}\ \bibnamefont {Martin-Delgado}},\
  }\href {\doibase 10.1103/PhysRevLett.120.017001} {\bibfield  {journal}
  {\bibinfo  {journal} {Phys. Rev. Lett.}\ }\textbf {\bibinfo {volume} {120}},\
  \bibinfo {pages} {017001} (\bibinfo {year} {2018})}\BibitemShut {NoStop}%
\bibitem [{\citenamefont {DeGottardi}\ \emph {et~al.}(2013)\citenamefont
  {DeGottardi}, \citenamefont {Thakurathi}, \citenamefont {Vishveshwara},\ and\
  \citenamefont {Sen}}]{diptiman1}%
  \BibitemOpen
  \bibfield  {author} {\bibinfo {author} {\bibfnamefont {W.}~\bibnamefont
  {DeGottardi}}, \bibinfo {author} {\bibfnamefont {M.}~\bibnamefont
  {Thakurathi}}, \bibinfo {author} {\bibfnamefont {S.}~\bibnamefont
  {Vishveshwara}}, \ and\ \bibinfo {author} {\bibfnamefont {D.}~\bibnamefont
  {Sen}},\ }\href {\doibase 10.1103/PhysRevB.88.165111} {\bibfield  {journal}
  {\bibinfo  {journal} {Phys. Rev. B}\ }\textbf {\bibinfo {volume} {88}},\
  \bibinfo {pages} {165111} (\bibinfo {year} {2013})}\BibitemShut {NoStop}%
\bibitem [{\citenamefont {Anderson}(1958)}]{AndersonLocalization}%
  \BibitemOpen
  \bibfield  {author} {\bibinfo {author} {\bibfnamefont {P.~W.}\ \bibnamefont
  {Anderson}},\ }\href {\doibase 10.1103/PhysRev.109.1492} {\bibfield
  {journal} {\bibinfo  {journal} {Phys. Rev.}\ }\textbf {\bibinfo {volume}
  {109}},\ \bibinfo {pages} {1492} (\bibinfo {year} {1958})}\BibitemShut
  {NoStop}%
\bibitem [{\citenamefont {Groth}\ \emph {et~al.}(2009)\citenamefont {Groth},
  \citenamefont {Wimmer}, \citenamefont {Akhmerov}, \citenamefont
  {Tworzyd\l{}o},\ and\ \citenamefont {Beenakker}}]{TAI1}%
  \BibitemOpen
  \bibfield  {author} {\bibinfo {author} {\bibfnamefont {C.~W.}\ \bibnamefont
  {Groth}}, \bibinfo {author} {\bibfnamefont {M.}~\bibnamefont {Wimmer}},
  \bibinfo {author} {\bibfnamefont {A.~R.}\ \bibnamefont {Akhmerov}}, \bibinfo
  {author} {\bibfnamefont {J.}~\bibnamefont {Tworzyd\l{}o}}, \ and\ \bibinfo
  {author} {\bibfnamefont {C.~W.~J.}\ \bibnamefont {Beenakker}},\ }\href
  {\doibase 10.1103/PhysRevLett.103.196805} {\bibfield  {journal} {\bibinfo
  {journal} {Phys. Rev. Lett.}\ }\textbf {\bibinfo {volume} {103}},\ \bibinfo
  {pages} {196805} (\bibinfo {year} {2009})}\BibitemShut {NoStop}%
\bibitem [{\citenamefont {Li}\ \emph {et~al.}(2009)\citenamefont {Li},
  \citenamefont {Chu}, \citenamefont {Jain},\ and\ \citenamefont
  {Shen}}]{TAI2}%
  \BibitemOpen
  \bibfield  {author} {\bibinfo {author} {\bibfnamefont {J.}~\bibnamefont
  {Li}}, \bibinfo {author} {\bibfnamefont {R.-L.}\ \bibnamefont {Chu}},
  \bibinfo {author} {\bibfnamefont {J.~K.}\ \bibnamefont {Jain}}, \ and\
  \bibinfo {author} {\bibfnamefont {S.-Q.}\ \bibnamefont {Shen}},\ }\href
  {\doibase 10.1103/PhysRevLett.102.136806} {\bibfield  {journal} {\bibinfo
  {journal} {Phys. Rev. Lett.}\ }\textbf {\bibinfo {volume} {102}},\ \bibinfo
  {pages} {136806} (\bibinfo {year} {2009})}\BibitemShut {NoStop}%
\bibitem [{\citenamefont {Schubert}\ \emph {et~al.}(2012)\citenamefont
  {Schubert}, \citenamefont {Fehske}, \citenamefont {Fritz},\ and\
  \citenamefont {Vojta}}]{TAI3}%
  \BibitemOpen
  \bibfield  {author} {\bibinfo {author} {\bibfnamefont {G.}~\bibnamefont
  {Schubert}}, \bibinfo {author} {\bibfnamefont {H.}~\bibnamefont {Fehske}},
  \bibinfo {author} {\bibfnamefont {L.}~\bibnamefont {Fritz}}, \ and\ \bibinfo
  {author} {\bibfnamefont {M.}~\bibnamefont {Vojta}},\ }\href {\doibase
  10.1103/PhysRevB.85.201105} {\bibfield  {journal} {\bibinfo  {journal} {Phys.
  Rev. B}\ }\textbf {\bibinfo {volume} {85}},\ \bibinfo {pages} {201105}
  (\bibinfo {year} {2012})}\BibitemShut {NoStop}%
\bibitem [{\citenamefont {Castro}\ \emph {et~al.}(2015)\citenamefont {Castro},
  \citenamefont {L\'opez-Sancho},\ and\ \citenamefont
  {Vozmediano}}]{Vozmediano}%
  \BibitemOpen
  \bibfield  {author} {\bibinfo {author} {\bibfnamefont {E.~V.}\ \bibnamefont
  {Castro}}, \bibinfo {author} {\bibfnamefont {M.~P.}\ \bibnamefont
  {L\'opez-Sancho}}, \ and\ \bibinfo {author} {\bibfnamefont {M.~A.~H.}\
  \bibnamefont {Vozmediano}},\ }\href {\doibase 10.1103/PhysRevB.92.085410}
  {\bibfield  {journal} {\bibinfo  {journal} {Phys. Rev. B}\ }\textbf {\bibinfo
  {volume} {92}},\ \bibinfo {pages} {085410} (\bibinfo {year}
  {2015})}\BibitemShut {NoStop}%
\bibitem [{\citenamefont {Rakovszky}\ and\ \citenamefont
  {Asboth}(2015)}]{Disorder-QW}%
  \BibitemOpen
  \bibfield  {author} {\bibinfo {author} {\bibfnamefont {T.}~\bibnamefont
  {Rakovszky}}\ and\ \bibinfo {author} {\bibfnamefont {J.~K.}\ \bibnamefont
  {Asboth}},\ }\href {\doibase 10.1103/PhysRevA.92.052311} {\bibfield
  {journal} {\bibinfo  {journal} {Phys. Rev. A}\ }\textbf {\bibinfo {volume}
  {92}},\ \bibinfo {pages} {052311} (\bibinfo {year} {2015})}\BibitemShut
  {NoStop}%
\bibitem [{\citenamefont {Altland}\ \emph {et~al.}(2015)\citenamefont
  {Altland}, \citenamefont {Bagrets},\ and\ \citenamefont
  {Kamenev}}]{Altland2015}%
  \BibitemOpen
  \bibfield  {author} {\bibinfo {author} {\bibfnamefont {A.}~\bibnamefont
  {Altland}}, \bibinfo {author} {\bibfnamefont {D.}~\bibnamefont {Bagrets}}, \
  and\ \bibinfo {author} {\bibfnamefont {A.}~\bibnamefont {Kamenev}},\ }\href
  {\doibase 10.1103/PhysRevB.91.085429} {\bibfield  {journal} {\bibinfo
  {journal} {Phys. Rev. B}\ }\textbf {\bibinfo {volume} {91}},\ \bibinfo
  {pages} {085429} (\bibinfo {year} {2015})}\BibitemShut {NoStop}%
\bibitem [{\citenamefont {Liu}\ and\ \citenamefont
  {Guo}(2018)}]{quasiperiodicSSH}%
  \BibitemOpen
  \bibfield  {author} {\bibinfo {author} {\bibfnamefont {T.}~\bibnamefont
  {Liu}}\ and\ \bibinfo {author} {\bibfnamefont {H.}~\bibnamefont {Guo}},\
  }\href {\doibase https://doi.org/10.1016/j.physleta.2018.09.023} {\bibfield
  {journal} {\bibinfo  {journal} {Physics Letters A}\ } (\bibinfo {year}
  {2018}),\ https://doi.org/10.1016/j.physleta.2018.09.023}\BibitemShut
  {NoStop}%
\bibitem [{\citenamefont {Dutta}\ \emph {et~al.}(2016)\citenamefont {Dutta},
  \citenamefont {Saha},\ and\ \citenamefont {Jayannavar}}]{dutta}%
  \BibitemOpen
  \bibfield  {author} {\bibinfo {author} {\bibfnamefont {P.}~\bibnamefont
  {Dutta}}, \bibinfo {author} {\bibfnamefont {A.}~\bibnamefont {Saha}}, \ and\
  \bibinfo {author} {\bibfnamefont {A.~M.}\ \bibnamefont {Jayannavar}},\ }\href
  {\doibase 10.1103/PhysRevB.94.195414} {\bibfield  {journal} {\bibinfo
  {journal} {Phys. Rev. B}\ }\textbf {\bibinfo {volume} {94}},\ \bibinfo
  {pages} {195414} (\bibinfo {year} {2016})}\BibitemShut {NoStop}%
\bibitem [{\citenamefont {Odagaki}(1980)}]{odagaki2}%
  \BibitemOpen
  \bibfield  {author} {\bibinfo {author} {\bibfnamefont {T.}~\bibnamefont
  {Odagaki}},\ }\href {\doibase https://doi.org/10.1016/0038-1098(80)91206-5}
  {\bibfield  {journal} {\bibinfo  {journal} {Solid State Communications}\
  }\textbf {\bibinfo {volume} {33}},\ \bibinfo {pages} {861 } (\bibinfo {year}
  {1980})}\BibitemShut {NoStop}%
\bibitem [{\citenamefont {Stone}\ and\ \citenamefont
  {Joannopoulos}(1981)}]{LogOffdiag1}%
  \BibitemOpen
  \bibfield  {author} {\bibinfo {author} {\bibfnamefont {A.~D.}\ \bibnamefont
  {Stone}}\ and\ \bibinfo {author} {\bibfnamefont {J.~D.}\ \bibnamefont
  {Joannopoulos}},\ }\href {\doibase 10.1103/PhysRevB.24.3592} {\bibfield
  {journal} {\bibinfo  {journal} {Phys. Rev. B}\ }\textbf {\bibinfo {volume}
  {24}},\ \bibinfo {pages} {3592} (\bibinfo {year} {1981})}\BibitemShut
  {NoStop}%
\bibitem [{\citenamefont {Soukoulis}\ and\ \citenamefont
  {Economou}(1981)}]{soukoulis81}%
  \BibitemOpen
  \bibfield  {author} {\bibinfo {author} {\bibfnamefont {C.~M.}\ \bibnamefont
  {Soukoulis}}\ and\ \bibinfo {author} {\bibfnamefont {E.~N.}\ \bibnamefont
  {Economou}},\ }\href {\doibase 10.1103/PhysRevB.24.5698} {\bibfield
  {journal} {\bibinfo  {journal} {Phys. Rev. B}\ }\textbf {\bibinfo {volume}
  {24}},\ \bibinfo {pages} {5698} (\bibinfo {year} {1981})}\BibitemShut
  {NoStop}%
\bibitem [{\citenamefont {Zak}(1989)}]{zakphase1989}%
  \BibitemOpen
  \bibfield  {author} {\bibinfo {author} {\bibfnamefont {J.}~\bibnamefont
  {Zak}},\ }\href {\doibase 10.1103/PhysRevLett.62.2747} {\bibfield  {journal}
  {\bibinfo  {journal} {Phys. Rev. Lett.}\ }\textbf {\bibinfo {volume} {62}},\
  \bibinfo {pages} {2747} (\bibinfo {year} {1989})}\BibitemShut {NoStop}%
\bibitem [{\citenamefont {Delplace}\ \emph {et~al.}(2011)\citenamefont
  {Delplace}, \citenamefont {Ullmo},\ and\ \citenamefont
  {Montambaux}}]{delplace2011}%
  \BibitemOpen
  \bibfield  {author} {\bibinfo {author} {\bibfnamefont {P.}~\bibnamefont
  {Delplace}}, \bibinfo {author} {\bibfnamefont {D.}~\bibnamefont {Ullmo}}, \
  and\ \bibinfo {author} {\bibfnamefont {G.}~\bibnamefont {Montambaux}},\
  }\href {\doibase 10.1103/PhysRevB.84.195452} {\bibfield  {journal} {\bibinfo
  {journal} {Phys. Rev. B}\ }\textbf {\bibinfo {volume} {84}},\ \bibinfo
  {pages} {195452} (\bibinfo {year} {2011})}\BibitemShut {NoStop}%
\bibitem [{\citenamefont {Asb{\'o}th}\ \emph {et~al.}(2016)\citenamefont
  {Asb{\'o}th}, \citenamefont {Oroszl{\'a}ny},\ and\ \citenamefont
  {P{\'a}lyi}}]{asboth2016book}%
  \BibitemOpen
  \bibfield  {author} {\bibinfo {author} {\bibfnamefont {J.}~\bibnamefont
  {Asb{\'o}th}}, \bibinfo {author} {\bibfnamefont {L.}~\bibnamefont
  {Oroszl{\'a}ny}}, \ and\ \bibinfo {author} {\bibfnamefont {A.}~\bibnamefont
  {P{\'a}lyi}},\ }\href {https://books.google.es/books?id=RWKhCwAAQBAJ} {\emph
  {\bibinfo {title} {A Short Course on Topological Insulators: Band Structure
  and Edge States in One and Two Dimensions}}},\ Lecture Notes in Physics\
  (\bibinfo  {publisher} {Springer International Publishing},\ \bibinfo {year}
  {2016})\BibitemShut {NoStop}%
\bibitem [{\citenamefont {Ryu}\ \emph {et~al.}(2010)\citenamefont {Ryu},
  \citenamefont {Schnyder}, \citenamefont {Furusaki},\ and\ \citenamefont
  {Ludwig}}]{tenfoldway}%
  \BibitemOpen
  \bibfield  {author} {\bibinfo {author} {\bibfnamefont {S.}~\bibnamefont
  {Ryu}}, \bibinfo {author} {\bibfnamefont {A.~P.}\ \bibnamefont {Schnyder}},
  \bibinfo {author} {\bibfnamefont {A.}~\bibnamefont {Furusaki}}, \ and\
  \bibinfo {author} {\bibfnamefont {A.~W.~W.}\ \bibnamefont {Ludwig}},\ }\href
  {http://stacks.iop.org/1367-2630/12/i=6/a=065010} {\bibfield  {journal}
  {\bibinfo  {journal} {New Journal of Physics}\ }\textbf {\bibinfo {volume}
  {12}},\ \bibinfo {pages} {065010} (\bibinfo {year} {2010})}\BibitemShut
  {NoStop}%
\bibitem [{\citenamefont {Chen}\ and\ \citenamefont {Chiou}()}]{longrangeW}%
  \BibitemOpen
  \bibfield  {author} {\bibinfo {author} {\bibfnamefont {B.-H.}\ \bibnamefont
  {Chen}}\ and\ \bibinfo {author} {\bibfnamefont {D.-W.}\ \bibnamefont
  {Chiou}},\ }\href@noop {} {\ }\BibitemShut {NoStop}%
\bibitem [{\citenamefont {Li}\ \emph {et~al.}(2014)\citenamefont {Li},
  \citenamefont {Xu},\ and\ \citenamefont {Chen}}]{phasediag2vec}%
  \BibitemOpen
  \bibfield  {author} {\bibinfo {author} {\bibfnamefont {L.}~\bibnamefont
  {Li}}, \bibinfo {author} {\bibfnamefont {Z.}~\bibnamefont {Xu}}, \ and\
  \bibinfo {author} {\bibfnamefont {S.}~\bibnamefont {Chen}},\ }\href {\doibase
  10.1103/PhysRevB.89.085111} {\bibfield  {journal} {\bibinfo  {journal} {Phys.
  Rev. B}\ }\textbf {\bibinfo {volume} {89}},\ \bibinfo {pages} {085111}
  (\bibinfo {year} {2014})}\BibitemShut {NoStop}%
\bibitem [{\citenamefont {Fleishman}\ and\ \citenamefont
  {Licciardello}(1977)}]{fluctuationFleishman}%
  \BibitemOpen
  \bibfield  {author} {\bibinfo {author} {\bibfnamefont {L.}~\bibnamefont
  {Fleishman}}\ and\ \bibinfo {author} {\bibfnamefont {D.~C.}\ \bibnamefont
  {Licciardello}},\ }\href {http://stacks.iop.org/0022-3719/10/i=6/a=003}
  {\bibfield  {journal} {\bibinfo  {journal} {Journal of Physics C: Solid State
  Physics}\ }\textbf {\bibinfo {volume} {10}},\ \bibinfo {pages} {L125}
  (\bibinfo {year} {1977})}\BibitemShut {NoStop}%
\bibitem [{\citenamefont {Cheraghchi}\ \emph {et~al.}(2005)\citenamefont
  {Cheraghchi}, \citenamefont {Fazeli},\ and\ \citenamefont
  {Esfarjani}}]{locdeloc}%
  \BibitemOpen
  \bibfield  {author} {\bibinfo {author} {\bibfnamefont {H.}~\bibnamefont
  {Cheraghchi}}, \bibinfo {author} {\bibfnamefont {S.~M.}\ \bibnamefont
  {Fazeli}}, \ and\ \bibinfo {author} {\bibfnamefont {K.}~\bibnamefont
  {Esfarjani}},\ }\href {\doibase 10.1103/PhysRevB.72.174207} {\bibfield
  {journal} {\bibinfo  {journal} {Phys. Rev. B}\ }\textbf {\bibinfo {volume}
  {72}},\ \bibinfo {pages} {174207} (\bibinfo {year} {2005})}\BibitemShut
  {NoStop}%
\bibitem [{\citenamefont {Zhou}\ and\ \citenamefont
  {Bhatt}(2003)}]{offdiaglongrange}%
  \BibitemOpen
  \bibfield  {author} {\bibinfo {author} {\bibfnamefont {C.}~\bibnamefont
  {Zhou}}\ and\ \bibinfo {author} {\bibfnamefont {R.~N.}\ \bibnamefont
  {Bhatt}},\ }\href {\doibase 10.1103/PhysRevB.68.045101} {\bibfield  {journal}
  {\bibinfo  {journal} {Phys. Rev. B}\ }\textbf {\bibinfo {volume} {68}},\
  \bibinfo {pages} {045101} (\bibinfo {year} {2003})}\BibitemShut {NoStop}%
\bibitem [{\citenamefont {Budoyo}(2008)}]{thesisConductanceRings}%
  \BibitemOpen
  \bibfield  {author} {\bibinfo {author} {\bibfnamefont {R.~P.}\ \bibnamefont
  {Budoyo}},\ }\emph {\bibinfo {title} {Conductance in Mesoscopic Rings}},\
  \href@noop {} {Ph.D. thesis},\ \bibinfo  {school} {Wesleyan University,
  Connecticut} (\bibinfo {year} {2008})\BibitemShut {NoStop}%
\bibitem [{\citenamefont {Inui}\ \emph {et~al.}(1994)\citenamefont {Inui},
  \citenamefont {Trugman},\ and\ \citenamefont {Abrahams}}]{sublattice1}%
  \BibitemOpen
  \bibfield  {author} {\bibinfo {author} {\bibfnamefont {M.}~\bibnamefont
  {Inui}}, \bibinfo {author} {\bibfnamefont {S.~A.}\ \bibnamefont {Trugman}}, \
  and\ \bibinfo {author} {\bibfnamefont {E.}~\bibnamefont {Abrahams}},\ }\href
  {\doibase 10.1103/PhysRevB.49.3190} {\bibfield  {journal} {\bibinfo
  {journal} {Phys. Rev. B}\ }\textbf {\bibinfo {volume} {49}},\ \bibinfo
  {pages} {3190} (\bibinfo {year} {1994})}\BibitemShut {NoStop}%
\bibitem [{\citenamefont {Breuer}\ and\ \citenamefont
  {Petruccione}(2007)}]{breuer2007book}%
  \BibitemOpen
  \bibfield  {author} {\bibinfo {author} {\bibfnamefont {H.}~\bibnamefont
  {Breuer}}\ and\ \bibinfo {author} {\bibfnamefont {F.}~\bibnamefont
  {Petruccione}},\ }\href {https://books.google.es/books?id=DkcJPwAACAAJ}
  {\emph {\bibinfo {title} {The Theory of Open Quantum Systems}}}\ (\bibinfo
  {publisher} {OUP Oxford},\ \bibinfo {address} {Great Clarendon Street, Oxford
  OX2 6DP},\ \bibinfo {year} {2007})\BibitemShut {NoStop}%
\bibitem [{\citenamefont {Benito}\ \emph {et~al.}(2016)\citenamefont {Benito},
  \citenamefont {Niklas}, \citenamefont {Platero},\ and\ \citenamefont
  {Kohler}}]{benito2016}%
  \BibitemOpen
  \bibfield  {author} {\bibinfo {author} {\bibfnamefont {M.}~\bibnamefont
  {Benito}}, \bibinfo {author} {\bibfnamefont {M.}~\bibnamefont {Niklas}},
  \bibinfo {author} {\bibfnamefont {G.}~\bibnamefont {Platero}}, \ and\
  \bibinfo {author} {\bibfnamefont {S.}~\bibnamefont {Kohler}},\ }\href
  {\doibase 10.1103/PhysRevB.93.115432} {\bibfield  {journal} {\bibinfo
  {journal} {Phys. Rev. B}\ }\textbf {\bibinfo {volume} {93}},\ \bibinfo
  {pages} {115432} (\bibinfo {year} {2016})}\BibitemShut {NoStop}%
\bibitem [{\citenamefont {Ruocco}\ and\ \citenamefont
  {G\'omez-Le\'on}(2017)}]{Ruocco2017}%
  \BibitemOpen
  \bibfield  {author} {\bibinfo {author} {\bibfnamefont {L.}~\bibnamefont
  {Ruocco}}\ and\ \bibinfo {author} {\bibfnamefont {A.}~\bibnamefont
  {G\'omez-Le\'on}},\ }\href {\doibase 10.1103/PhysRevB.95.064302} {\bibfield
  {journal} {\bibinfo  {journal} {Phys. Rev. B}\ }\textbf {\bibinfo {volume}
  {95}},\ \bibinfo {pages} {064302} (\bibinfo {year} {2017})}\BibitemShut
  {NoStop}%
\bibitem [{\citenamefont {Möller}\ \emph {et~al.}(2012)\citenamefont
  {Möller}, \citenamefont {Mukherjee}, \citenamefont {Adolphs}, \citenamefont
  {Marchand},\ and\ \citenamefont {Berciu}}]{moller2012}%
  \BibitemOpen
  \bibfield  {author} {\bibinfo {author} {\bibfnamefont {M.}~\bibnamefont
  {Möller}}, \bibinfo {author} {\bibfnamefont {A.}~\bibnamefont {Mukherjee}},
  \bibinfo {author} {\bibfnamefont {C.~P.~J.}\ \bibnamefont {Adolphs}},
  \bibinfo {author} {\bibfnamefont {D.~J.~J.}\ \bibnamefont {Marchand}}, \ and\
  \bibinfo {author} {\bibfnamefont {M.}~\bibnamefont {Berciu}},\ }\href
  {http://stacks.iop.org/1751-8121/45/i=11/a=115206} {\bibfield  {journal}
  {\bibinfo  {journal} {Journal of Physics A: Mathematical and Theoretical}\
  }\textbf {\bibinfo {volume} {45}},\ \bibinfo {pages} {115206} (\bibinfo
  {year} {2012})}\BibitemShut {NoStop}%
\bibitem [{\citenamefont {Zajac}\ \emph {et~al.}(2016)\citenamefont {Zajac},
  \citenamefont {Hazard}, \citenamefont {Mi}, \citenamefont {Nielsen},\ and\
  \citenamefont {Petta}}]{zajac2016}%
  \BibitemOpen
  \bibfield  {author} {\bibinfo {author} {\bibfnamefont {D.~M.}\ \bibnamefont
  {Zajac}}, \bibinfo {author} {\bibfnamefont {T.~M.}\ \bibnamefont {Hazard}},
  \bibinfo {author} {\bibfnamefont {X.}~\bibnamefont {Mi}}, \bibinfo {author}
  {\bibfnamefont {E.}~\bibnamefont {Nielsen}}, \ and\ \bibinfo {author}
  {\bibfnamefont {J.~R.}\ \bibnamefont {Petta}},\ }\href {\doibase
  10.1103/PhysRevApplied.6.054013} {\bibfield  {journal} {\bibinfo  {journal}
  {Phys. Rev. Applied}\ }\textbf {\bibinfo {volume} {6}},\ \bibinfo {pages}
  {054013} (\bibinfo {year} {2016})}\BibitemShut {NoStop}%
\bibitem [{\citenamefont {Nevado}\ \emph {et~al.}(2017)\citenamefont {Nevado},
  \citenamefont {Fern\'andez-Lorenzo},\ and\ \citenamefont
  {Porras}}]{nevado2017}%
  \BibitemOpen
  \bibfield  {author} {\bibinfo {author} {\bibfnamefont {P.}~\bibnamefont
  {Nevado}}, \bibinfo {author} {\bibfnamefont {S.}~\bibnamefont
  {Fern\'andez-Lorenzo}}, \ and\ \bibinfo {author} {\bibfnamefont
  {D.}~\bibnamefont {Porras}},\ }\href {\doibase
  10.1103/PhysRevLett.119.210401} {\bibfield  {journal} {\bibinfo  {journal}
  {Phys. Rev. Lett.}\ }\textbf {\bibinfo {volume} {119}},\ \bibinfo {pages}
  {210401} (\bibinfo {year} {2017})}\BibitemShut {NoStop}%
\bibitem [{\citenamefont {Meier}\ \emph {et~al.}()\citenamefont {Meier},
  \citenamefont {Fangzhao}, \citenamefont {Dauphin}, \citenamefont {Maffei},
  \citenamefont {Massignan}, \citenamefont {Hughes},\ and\ \citenamefont
  {Bryce}}]{sshoffdiagexp}%
  \BibitemOpen
  \bibfield  {author} {\bibinfo {author} {\bibfnamefont {E.~J.}\ \bibnamefont
  {Meier}}, \bibinfo {author} {\bibfnamefont {A.~A.}\ \bibnamefont {Fangzhao}},
  \bibinfo {author} {\bibfnamefont {A.}~\bibnamefont {Dauphin}}, \bibinfo
  {author} {\bibfnamefont {M.}~\bibnamefont {Maffei}}, \bibinfo {author}
  {\bibfnamefont {P.}~\bibnamefont {Massignan}}, \bibinfo {author}
  {\bibfnamefont {T.~L.}\ \bibnamefont {Hughes}}, \ and\ \bibinfo {author}
  {\bibfnamefont {G.}~\bibnamefont {Bryce}},\ }\href@noop {} {\bibinfo
  {journal} {arXiv:1802.02109}\ }\BibitemShut {NoStop}%
\bibitem [{\citenamefont {Sanchez-Palencia}\ and\ \citenamefont
  {Lewenstein}(2010)}]{disordercoldatoms}%
  \BibitemOpen
\bibfield  {journal} {  }\bibfield  {author} {\bibinfo {author} {\bibfnamefont
  {L.}~\bibnamefont {Sanchez-Palencia}}\ and\ \bibinfo {author} {\bibfnamefont
  {M.}~\bibnamefont {Lewenstein}},\ }\href {\doibase
  https://doi.org/10.1038/nphys1507} {\bibfield  {journal} {\bibinfo  {journal}
  {Nature Physics}\ }\textbf {\bibinfo {volume} {6}},\ \bibinfo {pages}
  {87–95} (\bibinfo {year} {2010})}\BibitemShut {NoStop}%
\end{thebibliography}%

\end{document}